\documentclass[a4paper]{article}
\usepackage{ASVspoof2021}
\usepackage{epsfig,amssymb,amsmath}

\usepackage{todonotes}
\usepackage{booktabs}
\usepackage{float}
\usepackage{colortbl}
\PassOptionsToPackage{hyphens}{url}\usepackage{hyperref}

\usepackage[position=top]{subfig}
\ninept
\usepackage{url}

\title{Speech is Silver, Silence is Golden: \\ 
What do ASVspoof-trained Models Really Learn?
}

\makeatletter
\def\name#1{\gdef\@name{#1\\}}
\makeatother
\name{{\em Nicolas M. Müller$^1$, Franziska Dieckmann$^2$, Pavel Czempin$^2$}, \\
      {\em Roman U. Canals$^2$, Konstantin Böttinger$^1$ and  Jennifer Williams$^3$ }}

\makeatletter
\def\address#1{\gdef\@address{#1\\}}
\makeatother
\address{
$^1$Fraunhofer Institute for Applied and Integrated Security AISEC, Germany\\
$^2$Technical University of Munich, Germany\\
$^3$The University of Edinburgh, UK\\
{\small \tt nicolas.mueller@aisec.fraunhofer.de} \\
}

\begin{document}
\maketitle

\begin{abstract}
We present our analysis of a significant data artifact in the official 2019/2021 ASVspoof Challenge Dataset. 
We identify an uneven distribution of silence duration in the training and test splits, which tends to correlate with the target prediction label. 
Bonafide instances tend to have significantly longer leading and trailing silences than spoofed instances.
In this paper, we explore this phenomenon and its impact in depth.

% We compare several types of models trained on %a) 
% only the duration of the %\emph{leading} silence and b) only on the duration of 
% \emph{leading} and \emph{trailing} silence. 
We conduct several experiments where we train and evaluate models both with and without trimming the silence in the ASVspoof 2019 data.
Results show that models trained on only the duration of the leading silence perform suspiciously well, and achieve up to 85\% accuracy and an equal error rate (EER) of $15.1\%$.
At the same time,  
we observe that even for established antispoofing models such as RawNet2, removing silence during training leads to comparatively worse performance. In that case, EER increases from $3.6\%$ (with silence) to $15.5\%$ (trimmed silence).
Our findings suggest that previous work may, in part, have inadvertently learned the spoof/bonafide distinction by relying on the duration of silence as it appears in the official challenge dataset.
We discuss the potential consequences that this has for interpreting system scores in the challenge and how the ASV community may further consider this issue. 
%Consequently, it could mean that spoofing detection may not be as advanced as previous high-scores have led to believe.
% We proceed to train competitive state-of-the-art models on the datasets, and compare their performance with and without trimming of the leading silence.
% On average, the performance of these models drops by N percent, hinting again that the leading silence may contain inform
%We hope that by sharing these results, the ASV community can further evaluate this phenomenon.
\end{abstract}


% \fi

\section{Introduction}
%The ASVspoof 2019 database \cite{asvspoof2019database} is the most established dataset for automatic speaker verification.
Fake recordings of human speech have the potential to cause significant social and economic damage.
So-called deepfakes, i.e. human speech synthesized with a neural network, have already been used to scam a CEO for 243.000\$~\cite{AVoiceDe85:online}. Additionally, the potential for slander, misinformation, and fake news is enormous.
Thus, there is a need for automatic detection and verification of human speech.

The ASVspoof 2019 Challenge\footnote{\url{https://www.asvspoof.org/}} Dataset is the most established dataset for training and benchmarking systems designed for the detection of spoofed audio and audio deepfakes~\cite{asvspoof2019database}.
Overall, the dataset presents two related tasks: logical access (LA) and physical access (PA). All of the data is based on the VCTK corpus \cite{vctk}. 
In this paper, we focus on the data that is used for the LA task which is concerned with the detection of spoofed speech that was generated from voice conversion (VC), text-to-speech (TTS) synthesis, and vocoder copy-synthesis.
% In previous challenges, such as the 2019 challenge, the LA task was shown to be relatively easier compared to the PA task. This is due in part to the upper-bound quality that can be achieved by current speech synthesis methodology. 
% As long as speech synthesis techniques generate speech that is of sufficiently poor quality, then detecting the speech as a deepfake is feasible.
The ASVspoof 2019 dataset~\cite{wang2020asvspoof} includes spoofs created by a broad range of speech synthesis techniques and includes state-of-the-art models such as
% However, there are a handful of speech synthesis methods that generate speech with quality high enough that it is difficult to distinguish between natural speech and audio deepfakes such as
WaveNet~\cite{vanwavenet} and the top-performing neural network VC system from the 2018 Voice Conversion Challenge~\cite{Kinnunen2018, lorenzo2018voice}.

% It is important to be able to consider how different speech synthesis methods contribute to deepfake detection on the LA task. For example, if a particular type of speech synthesis method is very good, this would make deepfake detection more difficult, and a stronger ASVspoof countermeasure would need to be developed. 
The task of detecting audio deepfakes implies the need to look for features of the data that correspond to the ground-truth labels (bonafide vs. spoofed). Ordinarily, any discovered artifacts would correspond with the nature of a deepfake such as a noisy glitch, phase mismatch, reverberation, or loss of intelligibility~\cite{wu2012detecting, sahidullah2015comparison}. Likewise, data artifacts that are not related to deepfakes would be counterproductive for research and development purposes. 

During the process of training competitive models for deepfake detection, we observed a data artifact that is of interest to the ASVspoof community. 
%When training a model for our own submission to the ASVspoof 2021 challenge,
We noticed an irregularity regarding how silences are distributed in the dataset. 
%uneven distribution of durations of silences in the data.
This could be problematic since a model may learn to only, or at least partially, base its decision on the duration of the silence. Such instances of dataset artifacts are not unknown in machine learning as with the issue found in the Pascal VOC 2007 dataset, where all images of horses also contained a specific watermark \cite{lapuschkin2019unmasking}.

% moved this to related work

Any data-driven learner, such as a neural network, will pick up on this uneven distribution of silences in the ASVspoof data.
This raises the question: have previous methods learned to discriminate audio deepfakes solely, or at least partially, based on the duration of the leading and trailing silences? Our main contribution in this paper seeks to answer this question. In addition, we examine the performance of a model trained using the duration of silence as the only feature. We report how performance deteriorates with strong baselines from published work when the silence in training data has been trimmed. And we discuss the implications that this has for the larger ASVspoof community.

\begin{figure}
    \centering
    \caption{The duration of the leading + trailing silence per attack ID in the LA part of ASVspoof 2019. The blue bar visualizes the average silence duration in seconds for bonafide data, while the red shows the different attacks with their attack ID on the x-axis. The error bars highlight the standard deviation and the red horizontal line displays the average silence duration over all malicious attacks.}
    \subfloat[\centering The average silence duration in seconds per audio file for the data split `train'.]{
        \includegraphics[width=0.45\textwidth]{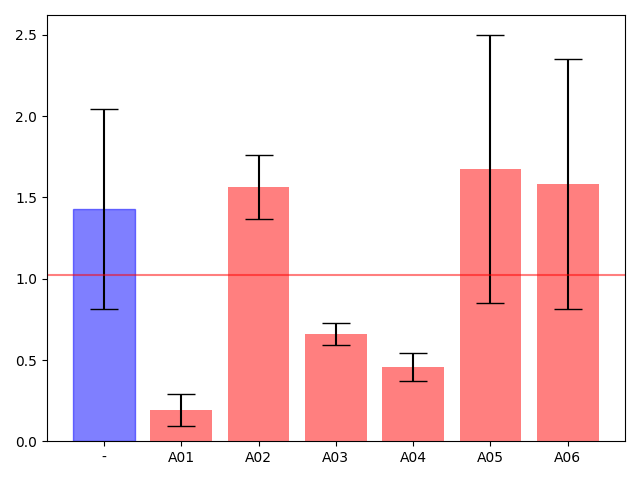}
        \label{fig:avg_sil_train}
    }
    \newline
    \subfloat[\centering This plot shows the average silence duration in seconds per audio file for the data split `dev'.]{
        \includegraphics[width=0.45\textwidth]{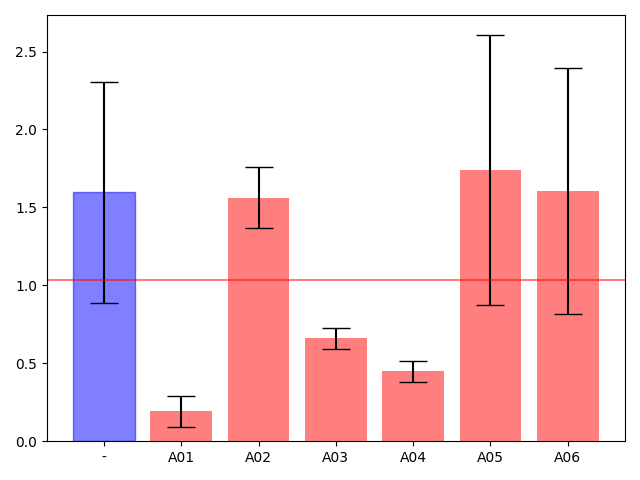}
        \label{fig:avg_sil_dev}
    }
    \newline
    \subfloat[\centering This plot shows the average silence duration in seconds per audio file for the data split `eval'.]{
        \includegraphics[width=0.45\textwidth]{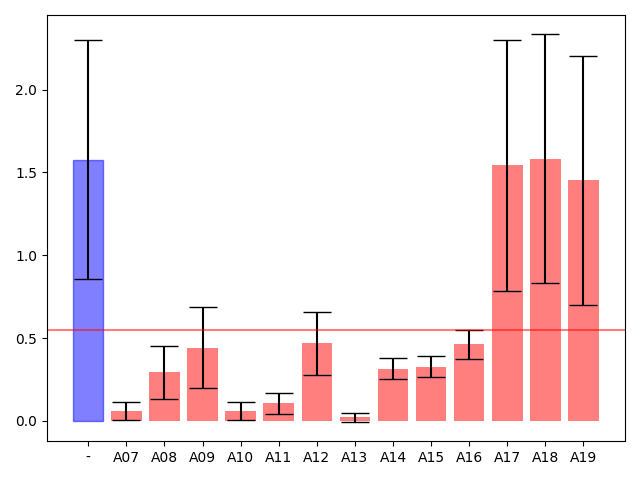}
        \label{fig:avg_sil_eval}
    }
    \label{fig:avg_sil}
\end{figure}

\section{Logical Access (LA) Dataset Description}
The dataset consists of pairs "speech, label" where speech is a .flac recording of human speech, either synthesized (label=spoof) or authentic (label=bonafide).
Already, several competitions have been organized, which all base their challenges on this dataset \cite{asvspoof2019challenge, asvspoof2021challenge}, and a significant number of related work has been published which uses ASVspoof as their baseline \cite{chintha2020recurrent, alzantot2019deep, chettri2019ensemble, wang2020densely}.
More details about the dataset are detailed in \cite{wang2020asvspoof}.

We observe an uneven distribution of the duration of silences in the ASVspoof 2019 dataset.
Figure \ref{fig:avg_sil} visualizes this for each of the three data splits `train', `dev', `eval'. 
The duration of the silence in the LA data seems to be a very informative feature to tell apart the bonafide (blue) and spoof (red) samples.
The bonafide samples have much longer silences than many of the attacks.
This is reasonable since the attacks are created by text-to-speech (TTS) systems, which are usually designed such that there is no leading or trailing silence in the output.
In fact, silences are usually trimmed from TTS training data, since arbitrary silences introduce uncertainty into the TTS training process.

%\section{Classifying based on leading silence}
\section{Model Descriptions}
In order to evaluate this phenomenon further, we implement the following models, which we use to empirically determine the amount of information contained in the dataset's silence.

\subsection{Our Baselines}\label{baselines}
First, we implement a random baseline, which has no trainable parameter and scores new samples with a random scalar $\lambda \in [0, ..., 1]$. We expect this model to yield an EER of $50\%$ on any given dataset or task.
Additionally, we implement a fully connected neural network (FCNN) with two hidden layers of size 128, ReLU activation, 10\% Dropout, and a single output neuron with sigmoid activation. This simple model serves as a baseline for more advanced learners.

\subsection{Established Models}
We implement three models, a Deep Residual Network, an LSTM, and a CNN-GRU model. These models are more powerful than the previously employed linear model and achieve an EER of up to $5.87$\%. Note that we train only on the `train' split, not on the `dev' split. %For more details, see the Appendix.
Note that all models except `RawNet2' accommodate variable-length input.

%In this section, we shortly describe the models used in Section~\ref{s:res2}.

\textbf{ResNet}. This model is a deep residual pre-activation network similar to~\cite{alzantot2019deep}. It consists of a stack of four residual layers, each of which consists of two stacks of batch-norm, leaky-relu, and a two-dimensional convolutional network.
We use a kernel size of $(3, 3)$, stride of $(1, 1)$ and padding of $(1, 1)$.
Since these configurations do not shrink the spatial dimensions, each ResNet block employs AveragePooling along the feature dimension, preserving time dimension (i.e. kernel size of $(3, 1)$).
After the last ResNet block,
we reshape the output to shape $(B, L, F)$ where $B$ is the batch-size, $L$ is the length of the time dimension, and $F$ is the flattened feature vector.
Finally, two dense layers compress this into $(B, L, 1)$, which is aggregated via \emph{mean} along the time axis and yields a single logit $\lambda \in \mathbb{R}$ per element in the mini-batch.

\textbf{CNN}. This model closely follows the architecture presented in \cite{wang2018style}, and consists of a stack of convolutional layers with (3,3) kernels, each with batch-norm and ReLU activation, followed by a Gated Recurrent Network. The CNN output is reshaped to conserve the time dimension (i.e. aggregating channel and feature dimension) and fed into a three-layer GRU, whose outputs are fed into two linear layers with 50\% dropout. The final linear layer has a single output neuron, followed by sigmoid activation. These linear layers classify each feature-frame separately, and the result is aggregated via mean over the time dimension and finally activated via the sigmoid function.

\textbf{LSTM}. This model follows \cite{jia2018transfer}, and consists of a three-layered, bi-directional LSTM with 10\% percent dropout and 256 hidden neurons.
The output of the LSTM is fed into a linear projection layer, which has a single output neuron.
We take the mean overall outputs along the time dimension, which yields a single logit $\lambda \in \mathbb{R}$. Finally, we apply sigmoid activation to match the targets $\in [0, 1]$.

\textbf{RawNet2}. This model~\cite{tak2021end} is a state-of-the-art convolutional/residual neural network using raw features as input.
This means that the audio is not transformed to spectogram, but that the waveform (16000 samples per second) is supplied to the model directly, which then employs stacks of convolutional layers to extract higher-level features.
This model is an official baseline for the ASVspoof 2021 challenge. We use the source code provided by the authors.

\section{Experimental Design}

\subsection{Training setup}\label{ss:setup}
We train our models on the `train' split, and evaluate their performance both on the `dev' and `eval' split.
We implement the models in Python 3.8 using PyTorch and run the experiments on an Nvidia A100 GPU.
We train the models for $50$ epochs each with a learning rate of $0.001$, using the Adam Optimizer~\cite{kingma2014adam} with a weight-decay of $1e^{-6}$.
No early stopping is employed.
We standard-normalize our features (subtracting the mean and dividing by the standard deviation) and use binary cross-entropy loss.

\subsection{Baseline Silence Features}
In order to evaluate the impact of the leading silence, we train the FCNN described in Section~\ref{baselines} on the ASVspoof 2019 dataset, but extract only a \textbf{single} feature: The duration of the leading silence in seconds. If there really is information contained in just the \emph{duration} of the silence, we expect the FCNN to outperform the random baseline, i.e. obtain EER less than $50\%$. Section~\ref{s:res1} presents the results.
% The labels are encoded as $0$ (bonafide) and $1$ (spoof).

\begin{table}[t]
    \vspace{2mm}
    \caption{The results of training a dense neural network on only the \emph{duration} of leading silence, compared against a random baseline.
    The network is able to significantly outperform the random baseline (see test EER, highlighted in blue). Results for the 2021 data (Deepfake and Logical Access) are copied verbatim from the submission platform (\emph{progress} phase).}
    \centerline{
        \begin{tabular}{llllll}
\toprule
Model & Dev EER & Eval EER &LA21 &DF21 \\
\midrule
     FCNN &      31.09±3.8 &        \cellcolor{blue!25}15.12±0.1 & 19.93  & 17.42 \\
    Random &       50.0±0.0 &        50.0±0.0 & - & - \\
\bottomrule
\end{tabular}

% \begin{tabular}{llllll}
% \toprule
% Model & Train EER & Dev EER & Eval EER &LA21 &DF21 \\
% \midrule
%      FCNN & 34.17±3.3 &      31.09±3.8 &        \cellcolor{blue!25}15.12±0.1 & 19.93  & 17.42 \\
%     Rand. & 50.0±0.0 &      50.0±0.0 &        50.0±0.0 & - & - \\
% \bottomrule
% \end{tabular}
    }
    \label{tab:res}
\end{table}

\subsection{Silence Trimming and CQT-Features}\label{s:exp2}
Additionally, we use three stronger models, a Deep Residual Network (ResNet), an LSTM and a CNN-GRU model (c.f. Section~\ref{baselines}) in combination with CQT features~\cite{brown1991calculation}, and train them with and without access to the audio silence.
More specifically, we employ two different pre-processing techniques:
\begin{itemize}
    \item \textbf{Time-wise subselection:} The first pre-processing technique is dubbed 'time-wise subselection'.
    If enabled ($t > 0$), it returns a random subsection of the audio of duration $t$. For each epoch, a new random selection is returned. If $t = -1$, this pre-processing technique is disabled and the audio remains unchanged.
    We test two settings: $t = 2.4s$ and $t = -1$.
    For $t = 2.4s$, a randomly chosen slice of duration  $150$ (since $150*0.016s = 2.4s$) from the audio is returned (where the hop size is $0.016$ seconds).
    % A subselection of $-1$ returns the complete feature of shape $L, F$ (where $L$ is the time dimension and $F$ the feature dimension, for example the number of mel bins). 
    % A subselection of $150$ returns a random section of shape $150, F$, i.e. returns a randomly chosen chunk of duration $150*0.016s = 2.4s$ from the audio (where the hop size is $0.016$).
    This has, among others, the effect that the model does not have consistent access to the audio silence.
    \item \textbf{Trim Silence}. The second pre-processing technique trims, if enabled, the silence of all audios before returning them to the model.
    We use the librosa \cite{librosa} library for this and employ $librosa.effects.trim$ with a \emph{ref\_db} of $40$ as a threshold to get the duration of the leading silence.
\end{itemize}

We train these models with four possible combinations of the two data augmentation techniques (time-wise subselection, and silence trimming). These combinations yielded twelve configurations.
We run each of these two times, resulting in training 24 models in total.
We expect to see a deterioration in EER if indeed the duration of the silence leaks significant information w.r.t. the target label.
Results are presented in Section~\ref{s:res2}.
% The results are averaged, and we report the results plus standard deviation in Table~\ref{tab:res2}.
% For these models, the augmentation was done to the train and development splits, but not the evaluation split.

In another version of this experiment, we repeat the previous trial, but remove the silence only from the evaluation data. More specifically, we train a model on the unmodified `train' split of ASVspoof 2019 with full access to the original audio, i.e. \emph{with} access to the silences. We then evaluate the model on two versions of the `eval' split: First, we evaluate it on the unchanged evaluation data. Second, we test it on the evaluation data wherein the leading and trailing silence had been truncated.
Results are presented in Section~\ref{s:res2}.

\subsection{Silence Trimming and RawNet2}\label{s:exp3}
Additionally, we integrated RawNet2~\cite{tak2021end}, one of the official ASVspoof2021 baselines, into our test setup.
Since it uses \emph{feature subselection} by design ($t = 4s$), we skip the configuration $t=-1$, but evaluate only the impact of silence trimming.
We train the model for $50$ epochs and three times for each configuration and present the results in Section~\ref{s:res3}.
% \begin{itemize}
%     \item \textbf{Trim Silence} is set to True and False. This time, we only trim silence in the training set, but leave silence in the eval set.
%     % \item \textbf{Subselection} is set to either $t=4s$ or $t=10s$. The rational behind $t=10s$ is to mimic $t=-1$, since such a large window will surely capture the audio silence, if present.
% \end{itemize}

\section{Results and Analysis}

\subsection{Baseline Results}\label{s:res1}
Table~\ref{tab:res} shows the results when training the FCNN only on the \emph{duration} of the leading silence.
The data is averaged over four individual runs, and shown with standard deviation.
Our very simple model substantially outperforms the random baseline, and achieves a respectable $15.12\%$ EER on the test set (vs. $50\%$ EER of the random baseline), using only a single feature as input: the duration of the leading silence (i.e. only a single, scalar input).
Observe how the EER differs between the `dev' and `eval' dataset ($31.09\%$ vs $15.12\%$).
This is because the `eval' dataset has a larger number of attacks where the duration of the silence clearly indicates the label (A07-A16), c.f. Figure~\ref{fig:avg_sil_eval}.
Note that this phenomenon occurs also for the 2021 dataset (both for the 'LA' and 'deepfake' dataset), where the FCNN trained only on the duration of the audio silence significantly beats the 50\% EER threshold (we obtain 19.93\% and 17.42\% EER during the \emph{progress} phase% on July 5th, 2021
, c.f. Table~\ref{tab:res}).
% This is due to the `eval' dataset's instance having a more pronounced leading silence than those of the `train' and `dev' split (c.f. Figure~\ref{fig:avg_sil}).

%\subsection{Discussion}
% In the previous section, we show that a very basic DNN can obtain a EER of $0.15$ on the test set using only a single feature - the duration of the leading silence of the audio files.
More powerful deep learning models which use MFCC \cite{mfcc, davis1980comparison} or CQT \cite{brown1991calculation} features achieve around 4.0-7.0\% EER, c.f. Section~\ref{s:res2} and \cite{alzantot2019deep}.
While this is a significant improvement compared to an EER of 15.12\%, 
we suspect that these deep learning models use the duration of the leading and trailing silence to discriminate the audio files nevertheless.

\subsection{Results for Silence Trimming and CQT-Features}\label{s:res2}

This section presents the results of the experiment described in ~\ref{s:exp2}, where we train models from Section~\ref{baselines} both with and without access to the audio silence.
Table~\ref{tab:res2} presents the individual results.

Note that there is a single configuration of pre-processing steps where the model has the chance to consistently exploit the information contained in the duration of the leading silence: Subselection \textit{t = -1} and \textit{Trim Silence = False}.
For all models, this configuration considerably outperforms all other configurations:
The models net an EER of $5.87\%$, $7.48\%$, and $8.33\%$ (blue boxes) when they have reliable access to the leading silence.
In the other three scenarios, they never achieve more than $19.05\%$ EER.
This is less than what the very basic single-feature dense network in Section~\ref{s:res1} achieved.
This strongly hints at the fact that the models draw considerable, if not most of their discriminatory power from either the duration of the silence or the silence itself.

%\cellcolor{blue!25}
\begin{table}[]
    \centering
    \caption{Three Deep Learning Models are trained on the ASVspoof 2019 `train' split and evaluated on both `dev' and `eval', each with four different configurations of data augmentation. For each model, the test EER is lowest (blue box) when the data augmentation allows consistent access to the audio silence.}
    % \begin{tabular}{|l|r|l|l|l|l|}
% \hline
% Model&  Subsel. &  Trim Sil & Train EER & Dev EER & Eval EER \\
% \hline \hline
%       ResNet &             -1 &      False & 0.00±0.00 &      0.01±0.00 &       \cellcolor{blue!25}0.05±0.01 \\
%       ResNet &             -1 &       True & 0.01±0.00 &      0.09±0.07 &        0.27±0.03 \\
%       ResNet &            2.4s &      False & 0.01±0.00 &     0.02±0.01 &        0.25±0.01 \\
%       ResNet &            2.4s &       True & 0.01±0.00 &      0.02±0.01 &        0.29±0.01 \\
%   CNN &             -1 &      False & 0.00±0.00 &      0.00±0.00 &       \cellcolor{blue!25} 0.07±0.01 \\
%   CNN &             -1 &       True & 0.00±0.00 &      0.01±0.00 &        0.26±0.04 \\
%   CNN &            2.4s &      False & 0.00±0.00 &      0.01±0.00 &        0.20±0.01 \\
%   CNN &            2.4s &       True & 0.00±0.00 &      0.00±0.00 &        0.25±0.01 \\
%       LSTM &             -1 &      False & 0.00±0.00 &      0.01±0.01 &       \cellcolor{blue!25} 0.08±0.01 \\
%       LSTM &             -1 &       True & 0.01±0.00 &      0.06±0.01 &        0.27±0.01 \\
%       LSTM &            2.4s &      False & 0.01±0.00 &      0.04±0.00 &        0.19±0.01 \\
%       LSTM &            2.4s &       True & 0.01±0.00 &      0.06±0.01 &        0.26±0.00 \\
% \hline
% \end{tabular}

\begin{tabular}{|l|r|l|l|l|l|}
\hline
Model&  Trim Sil. & Subsel. &   Dev EER & Eval EER \\
\hline \hline
        ResNet &      False &             -1 &       7.08±1.5 &         \cellcolor{blue!25}5.87±1.8 \\
  ResNet &      False &            2.4s &       1.66±0.4 &        24.68±2.2 \\
  ResNet &       True &             -1 &       8.83±7.6 &        27.23±3.2 \\
  ResNet &       True &            2.4s &       1.83±0.1 &        29.13±1.6 \\
   CNN &      False &             -1 &       0.15±0.1 &         \cellcolor{blue!25}7.48±0.6 \\
   CNN &      False &            2.4s &       0.61±0.2 &        20.02±0.9 \\
   CNN &       True &             -1 &       0.68±0.4 &        26.27±3.5 \\
   CNN &       True &            2.4s &       0.34±0.1 &        25.47±0.9 \\
      LSTM &      False &             -1 &       0.84±0.5 &         \cellcolor{blue!25}8.33±1.2 \\
      LSTM &      False &            2.4s &       4.50±0.2 &        19.05±0.6 \\
      LSTM &       True &             -1 &       6.21±0.7 &        27.28±1.4 \\
      LSTM &       True &            2.4s &       6.44±1.2 &        25.62±0.4 \\
\hline
\end{tabular}

    \label{tab:res2}
\end{table}

We observe the same phenomenon when trimming silence only from the evaluation data (c.f. Table~\ref{tab:res3}):
When removing the silences from the eval dataset, EER increases from $7.35\%$ to over $35.32\%$.
This again strongly hints at the fact that the bulk of information w.r.t the target label is contained in the silence.

\subsection{Results for RawNet2 on ASVspoof 2021}\label{s:res3}
 Table~\ref{tab:res_rawnet} presents the results when evaluating RawNet2, one of the ASVspoof 2021 baselines.
We observe the same phenomenon as in previous experiments:
When the model has access to the audio silences during training (i.e. \emph{Trim Silence = False}), EER is low ($3.61\%$). When \emph{Trim Silence = True}, the EER of the 2019 `eval' data increases by a factor of five to an absolute value of $15.50\%$.

Additionally, we evaluate the models on the 2021 $DF$ and $LA$ datasets\footnote{
We evaluate the 2021 data only for exemplifying the silence issue and not because this is a system submission to the challenge.
}.
We observe the same trend when we submit the model's scores to the CodaLab submission platform during the \emph{progress} phase, c.f. Table~\ref{tab:res_rawnet}: For both the DF and the LA task, trimming the silence during training severely degrades model performance (EER increases from $6.28$\% to $20.79$\% for the deepfake, and $10.38$\% to $27.39$\% for the LA part).
We observe that suspiciously much information is contained in the audio silence, without which model performance degrades significantly.

% no access: 0.7247 (42)	26.62 (44)
% with access: 0.6253 (39)	18.45 (38)

%\cellcolor{blue!25}
\begin{table}[]
    \centering
    \caption{An LSTM model trained with leading + trailing silence, evaluated on the 2019 `eval' data, with and without trimming the silence from the evaluation data.
    Observe how removing the silences increases EER by 500\%.}
    \begin{tabular}{lrlll}
\toprule
Model Name &   Trim Sil. Eval & Eval EER \\
\midrule
      LSTM &             False &        7.35$\pm$ 1.3 \\
      LSTM &             True &        35.32$\pm$ 3.8 \\
\bottomrule
\end{tabular}

    \label{tab:res3}
\end{table}

%\cellcolor{blue!25}
\begin{table}[]
    \centering
    \caption{RawNet2\cite{tak2021end}, one of the baselines of the ASVspoof 2021 challenge, trained with and without access to the audio silence. Results averaged over three runs. Removing silence from the Training set prevents the model from using the duration of the silence in the eval set as cue. Consequently, 2019 `eval' EER deteriorates by about 500\%. The same phenomenon can be observed for the 2021 data from the Deepfake (DF) and Logical Access (LA) track (EER values copied as is from CodaLab submission website during the \emph{progress} phase on July 7th, 2021).}
    \begin{tabular}{ccccc}
\toprule
Model &   Trim Sil. & EER Eval-19 & DF-21 & LA-21\\
\midrule
%   RawNet2&             False &      \cellcolor{blue!25}0.036±0.005 & 6.43 & \\
%   RawNet2 &             True &      0.155±0.050 & 20.79 &\\
   RawNet2&             False &      \cellcolor{blue!25}3.61±1.2 &  \cellcolor{blue!25}6.28 &  \cellcolor{blue!25}10.38 \\
   RawNet2 &             True &      15.50±5.2 & 20.79 &27.39 \\
\bottomrule
\end{tabular}
~              
    \label{tab:res_rawnet}
\end{table}

\subsection{Duration of Silence as an Auxiliary Feature}
Finally, we try supplying the duration of leading + trailing silence as a feature to our models in order to see if this might improve performance.
More specifically, we supply the duration of the silence during both training and testing to our CNN, LSTM, and ResNet models, \emph{without} trimming the audio's silence, so the information on the duration of the silence is explicit but redundant.
However, we are unable to observe any improvements in EER.
We presume that the models are able to extract the duration of the silence easily enough on their own, such that there is no benefit in providing it manually.

% This causes models trained with this silence to use the duration of silence a and could potentially be exploited by an attacker to circumvent the spoof detection system.
% Additionally, for the ASVspoof 2017 replay detection challenge, \cite{chettri2020dataset} observe that the presence of certain attributes such as 'clicks', trailing and leading silence in the audio may hint at the target label.
% However, a thorough examination of the impact of the silence present in ASVspoof 2019 has, to the best of our knowledge, not been presented before. 
%Especially, the question as to what models really learn when training on ASVspoof2019 has not been considered w.r.t. the distribution of silences.

\section{Discussion: Silence in Spoofing-Detection}
We shortly discuss silence in spoofing-detection datasets such as ASVspoof.
Silence usually has (imperceptible) properties and characteristics, and includes faint background noises, artifacts from the microphone used, etc.
Thus, silence in a digital recording is represented by a series of non-zero scalar values of low magnitude.
It is reasonable to assume that a TTS system will not be able to fake silence perfectly, so silence can be legitimately used to classify audio recordings into spoof/bonafide.
Thus, in general, it is permissible for recordings in a dataset such as ASVspoof to include leading and trailing silence. 

What is problematic, however, is a very uneven distribution of silence. As we show in Figure~\ref{fig:avg_sil}, the mere \emph{duration} of silence strongly hints at the ground-truth label.
In this case, the model learns classification not due to the properties and characteristics of the silence itself, but due to its \emph{duration} (c.f. Table~\ref{tab:res}).
Thus, if silence is to be included in the dataset, it should be distributed evenly between the bonafide and spoofed instances.

\section{Related Work}
There is some previous work that examines the impact of silence in ASVspoof.
% The issue of silence padding in the ASVspoof 2019 dataset was first presented in \cite{chettri2020voice}. 
For example, \cite{chettri2019ensemble} find that the duration of silence in the PA part of ASVspoof 2019 differs between bonafide and spoof data. 
Additionally, \cite{chettri2020voice} performed an analysis of silence injection as well as silence padding. In particular, they found that silence padding correlated with bonafide vs. spoofed ground truth labels. Their work focused on how silence (both injected and original padding) affects the performance of systems such as Gaussian mixture models (GMMs) and support vector machines (SVMs). While the analysis is insightful, they did not re-evaluate top-performing systems from the challenge or examine silence as a standalone feature and that is what is addressed in this paper.
% \todo{need to double-check that this is the correct interpretation of the results from their work}

\section{Conclusion}
We have examined the data that was used in the Logical Access (LA) task of the ASVspoof 2019 Challenge with respect to the distribution of leading and trailing silences.
We demonstrated that simply the \emph{duration} of the silence provides the deep learning models with considerable information about the bonafide vs. spoof labels.
When the leading and trailing silences are removed, performance deteriorates for strong models that were published in related work, including ResNet, LSTM, and CNN-GRU.
Our experiments indicate that models trained on the original ASVspoof 2019 LA dataset may learn to classify bonafide vs. spoof speech audio primarily based on the duration of the silence. Understandably, this may not be what the ASVspoof Challange organizers had intended, and it could potentially inhibit progress in this very important research

Since a lot of related work uses ASVspoof as a benchmark, it may be necessary to take appropriate steps.
A simple solution may be to publish a second version of the dataset, where the leading silence has been removed via the Python librosa library or a forced aligner such as the Montreal Forced Aligner (MFA) \cite{mcauliffe2017montreal}. Another possible remedy is to add more training, development, and evaluation examples such that the duration of the leading silence is balanced between bonafide vs. spoof labels, as well as attack types.

We have shown that the presence/absence of leading and trailing silence in speech audio is sufficient to significantly change the EER performance of various types of models in the ASVspoof 2019 Challenge. Preliminary results indicate that this extends to the ASVspoof 2021 datasets. 
We suspect that techniques from previous work may learn to classify audio mostly based on the duration of the silence. 
%Since silence is not a feature that will generalize in real-world applications for spoofing attack detection, previously achieved high-scores should be re-evaluated.
This makes generalization complicated.
%  Since different DNNs and signal features will handle silence differently, the comparison between anti-spoofing techniques is weakened. Results from related work might not be reporting how well the systems perform on speech. And, because of this, systems that were lower-ranked in the previous challenges might actually be better at anti-spoofing.
 %because of how those system represents silence in their features and architectures.

\bibliographystyle{unsrt}
\bibliography{bibliography}

% \begin{table}[]
%     \centering
%     \input{tab/res4}
%     \caption{A CNN model trained on raw audio with leading silence, evaluated on the `eval' split of ASVspoof 2019 with and without leading silence.
%     Observe how removing the silences from the eval increases EER by 300\%.}
%     \label{tab:res4}
% \end{table}

\end{document}